\documentstyle[prd,aps,epsf]{revtex}

\newcommand{\be}{\begin{equation}}
\newcommand{\ee}{\end{equation}}
\newcommand{\bea}{\begin{eqnarray}}
\newcommand{\eea}{\end{eqnarray}}
\newcommand{\IR}{\bf R} 

\def\IZ{\relax\ifmmode\hbox{Z\kern-.4em Z}\else{Z\kern-.4em Z}\fi}
\newcommand{\IS}{{\bf S}}

\draft

\begin{document}
\twocolumn[\hsize\textwidth\columnwidth\hsize\csname
@twocolumnfalse\endcsname

\title{Speculative generalization of black hole uniqueness to higher dimensions}

\author{Barak Kol}

\address{School of Natural Sciences \\ Institute for Advanced Study \\
Einstein Drive, Princeton NJ 08540, USA\\
email barak@sns.ias.edu}

\maketitle

\begin{abstract}
A straightforward generalization of the celebrated uniqueness
theorem to dimensions greater than four was recently found to
fail in two pure gravity cases - the 5d rotating black ring and
the black string on $\IR^{3,1} \times \IS^1$. Two amendments are
suggested here (without proof) in order to rectify the situation.
The first is that in addition to specifying the mass and angular
momentum (and gauge charges) one needs to specify the horizon
topology as well. Secondly, the theorem may survive if applied
exclusively to stable solutions. Note that the latter is at odds
with the proposed stable but non-uniform string.
\end{abstract}


\vskip2pc]

\vskip1cm

\section{Introduction}

The celebrated uniqueness theorem states that 4d black holes with
regular horizons are determined uniquely by their mass $M$ and
angular momentum $J,\: |J|/M^2 \le 1$ in pure gravity and by
their electric (and possibly magnetic) charge when a Maxwell
field is present, with important contributions by Israel, Carter,
Hawking, Robinson, Mazur, Bunting and Masood Ul Alam and others.
See for example \cite{IsraelUnique,CarterUnique,RobinsonUnique}
and for a thorough reference list see the reviews
\cite{MazurReview,HeuslerReview,CarterReview,ChruscielReview} and
references therein.

Recently two cases were studied in which a straightforward
generalization of black hole uniqueness to more than four
dimensions fails. The first is the black-string to black-hole
transition which takes place on any compactification background
the simplest being $\IR^{3,1} \times \IS^1$ background -- $3+1$
extended dimensions and a fifth compact dimension of radius $L$,
which is denoted here by the $z$ coordinate.  Black objects with
Schwarzschild radius much smaller than $L$ are expected to
closely resemble a 5d Schwarzschild black hole (BH) with a nearly
round $\IS^3$ horizon topology, while very massive black objects,
will have a horizon size much larger than $L$ and will be
extended over the $z$ axis wrapping it completely. Such objects
will have $\IS^2 \times \IS^1$ horizon topology and will be
referred to as ``black strings''. One type of black string is the
4d Schwarzschild solution with the $z$ coordinate added in as a
spectator and it will be called here a ``uniform string'', while
another type of solution may be $z$ dependent, namely, a
``non-uniform string''. Gregory and Laflamme discovered an onset
of tachyonic instability \cite{GL1,GL2} when the uniform black
string becomes so thin that its Schwarzschild radius is of the
same order as $L$. Following the claimed existence of a stable
non-uniform solution \cite{HorowitzMaeda} Gubser performed a
higher order analysis of the Greogory-Laflamme (GL) critical
point and found that the non-uniform solution which must emerge
there (as the GL mode is marginally tachyonic) is unstable and its
mass is larger than the critical GL mass \cite{Gubser}.

\begin{figure}
\centerline{\epsfxsize=70mm\epsfbox{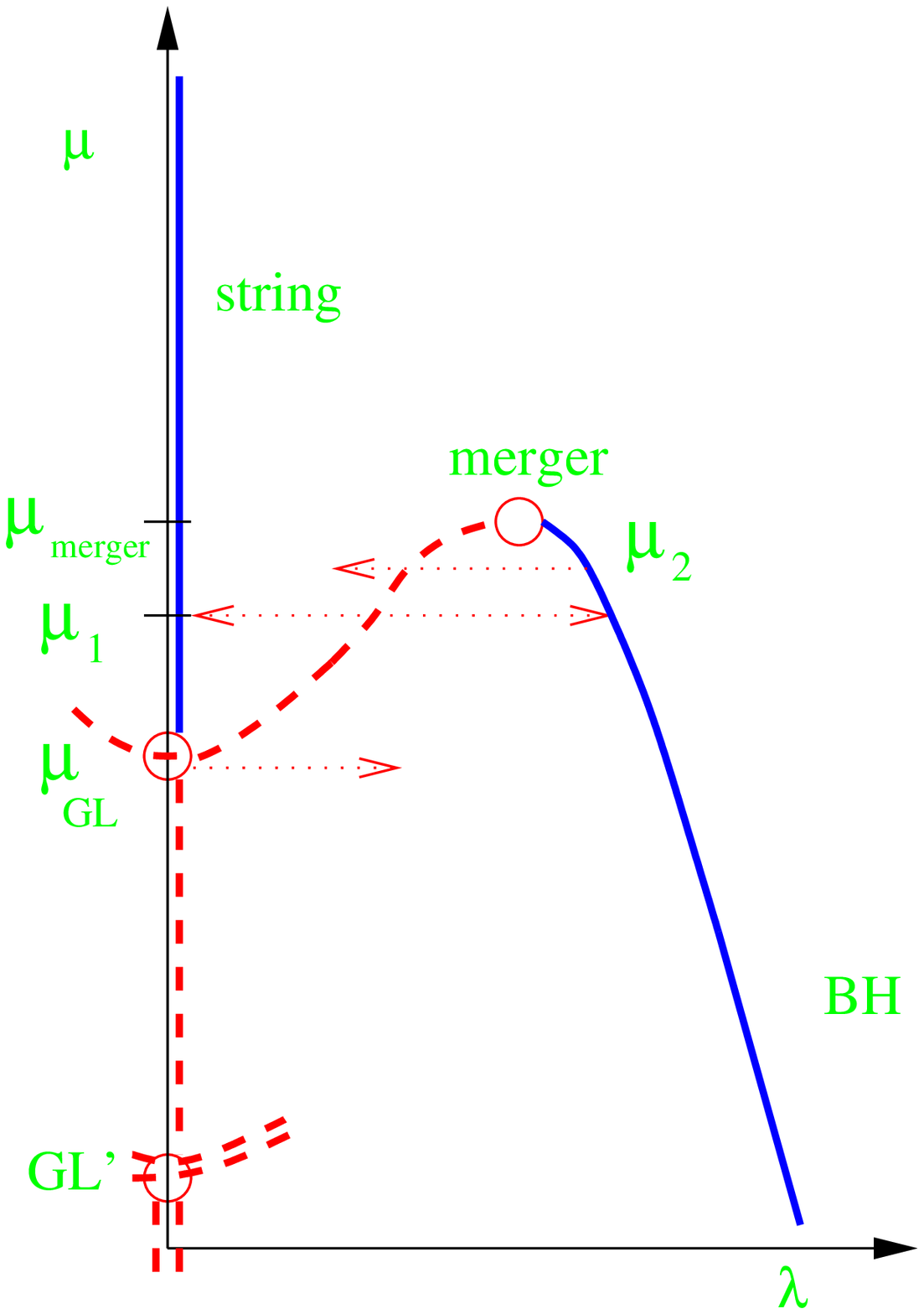}}
\medskip
\caption{A suggested phase diagram for the black-hole black-string
transition 
[12]. $\mu$ is the dimensionless parameter $G_5\, M/L^2$, and
$\lambda$ is an order parameter (measure of non-uniformity). In
the range $\mu_{GL} \le \mu \le \mu_{\mbox{merger}}$ (at least)
three solutions coexist. Solid lines are stable phases while
dashed ones are unstable. The solutions to the left of the merger
point are black strings while those to the right are black holes.
Dotted lines denote transitions - a first order transition at
$\mu_1$ and tachyonic decay from the two other points. The full
phase diagram should include further refinements in the area of
the merger transition.}
 \label{phasediag}
\end{figure}

Later an analysis of the phase diagram of the whole process,
supplemented by a local demonstration of topology change near the
cone over $\IS^2 \times \IS^2$ produced the following phase
diagram (figure \ref{phasediag})
\cite{BKTopologyChange,BKExplosive}. In this diagram it was not
possible to accommodate the stable non-uniform string without
creating some other serious problem. The violation of uniqueness
is apparent for a range of masses above the GL critical mass,
where three solutions are present: the uniform string, the 5d
black hole and an unstable (unlike the previous two) and
non-uniform string. Closer to the merger point probably there are
even more phases but with more unstable modes as well.

\begin{figure}
\centerline{\epsfxsize=60mm\epsfbox{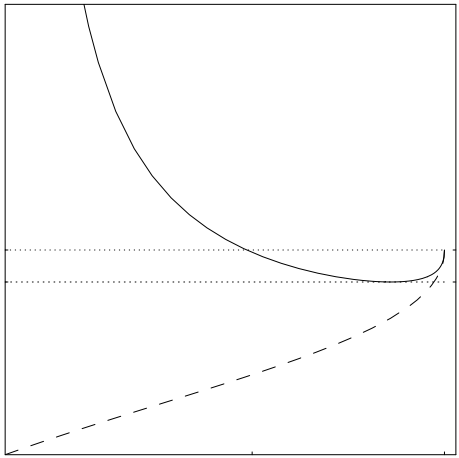}}
\medskip
\caption{Rotating solutions in 5d. $(27\, \pi/32\, G)\, J^2\,
/M^3$ as a function of some parameter $\nu$. The dashed line is
the rotating black hole of
[14] while the solid line is the newly discovered
 black ring 
[15]. In the range $J_1 \le J \le J_2$ marked by the dotted lines
three solutions coexist, denoted I, II and III from {\it right to
left}. Reproduced with permission from [15].}
 \label{BlackRing}
\end{figure}

The second example is the 5 dimensional rotating black object (in
uncompactified pure gravity). Here there are two kinds of
solutions, a black hole with horizon topology $\IS^3$
\cite{MyersPerry} and the recently discovered rotating ring with
horizon $\IS^2 \times \IS^1$ \cite{EmparanReallRing}. For given
mass $M$ the black hole exists for an angular momentum range $|J|
\le J_2$ for some $J_2$, while the ring exists for $J \ge J_1$
for some $J_1 < J_2$. As can be seen in figure (\ref{BlackRing})
uniqueness is violated in the range $J_1 < J < J_2$  (as noted by
the authors) where three solutions exist: two rings and one black
hole. The stability/ instability of either of these solutions is
not known at this time. Moreover, in 5d there are two parameters
of angular momentum $a_1,\, a_2$ and this picture emerged for
$a_2=0$ but it is not known yet how it generalizes to $a_2 \neq
0$.

Following \cite{EmparanReallRing} uniqueness in higher dimensions
was reexamined and the straightforward generalization was found
to hold for the static ($J=0$) uncompactified case and also when
some scalar fields are included
\cite{Hwang,Gibbons:2002bh,Gibbons:2002av,Gibbons:2002ju}. Other
relatively recent related works include \cite{Rogatko,Mars,Wells}.

Here a different approach will be taken, and rather than trying
to map the set of cases where straightforward uniqueness holds
and where it does not, I will speculate on an appropriate
generalization of uniqueness that may hold without exception,
hoping such a generalization exists since considering the
fundamental importance of the uniqueness theorem in 4d it would
be a shame if it simply disappeared in higher dimensions.

The uniqueness theorem can be thought as being made out of two
steps - the counting of parameters of black holes (``no  hair'')
and the impossibility of bifurcations or disconnected solution
families. The parameter counting is the fundamental physical
property, related to the amount of required boundary conditions,
and it is not violated in higher dimensions. Actually, once the
counting is done for small black holes with small angular
momenta, it is bound to hold always including for black holes in
the presence of a compactification or a cosmological constant.
Thus we are left to examine bifurcations and disconnected
components, which rely on a different principle for proof --
the  positive energy theorems.

Guided by the phenomenology, namely the two cases described above,
it is seen that bifurcations do occur, and two criteria suggest
themselves to me. The first is that the {\it horizon topology}
should be added to the list of (continuous) characterizing
parameters such as mass etc. Yet in both cases this is not enough
to uniquely specify a solution. Thus the second criterion of {\it
stability} is suggested. Actually this is better thought to be a
raised question, namely does uniqueness hold if solutions (with
identical horizon topologies) are limited to be stable? This
condition is physically attractive since unstable solutions
cannot be attained anyway (due to noise).

The other possibility is that there is no generalization of
uniqueness to higher dimensions and phase diagrams with several
phases are common without any simple constraints on the number or
type of phases.

\section{Horizon topology}

In 4 dimensions the only possible topology for a horizon is
$\IS^2$ (forgetting the light-like direction $\simeq \IR$)
\cite{HawkingS2Horizon}. However, the examples above show that in
higher dimensions the horizon topology is not unique. Since any
discussion of uniqueness of time independent solutions starts
with a dimensional reduction along Killing vectors and analysis
of the boundary conditions, where the boundary is composed of
asymptotic infinity and the event horizon, it is clear that the
horizon topology must be specified as one of these boundary
conditions.

One may ask what are the possible horizon topologies in higher
dimensions. From the examples it is seen that in 5d both $\IS^3$
and $\IS^2 \times \IS^1$ are possible. By considering toroidal
compactifications a generalized black string in $d$ dimensions
could have {\it any} $\IS^k \times {\bf T}^{d-2-k}$ {\it topology}
where ${\bf T}^n \simeq (\IS^1)^n$ is the $n$-dimensional torus,
and $k \ge 2$.

More generally we may consider compactification over an arbitrary
Ricci flat manifold, and consider the changes in topology of a
small black hole as its mass is being increased. The possible
topologies are not known, other than the obvious restriction that
they should be cobordant with the topology at asymptotic infinity
(and the cobordism type may be constrained in certain theories).
A possible guide for topologies could be the change in topology
of the equipotential lines of a (scalar) point source on the
compactification manifold as one goes away from the source. It
would be interesting to find a simple characterization of these
topologies, if one exists.

\section{Stability}

In both of the cases studied here horizon topology is not enough
to specify a solution in the parameter range of coexistence. Thus
I am led to wonder whether adding a stability requirement would
do. In the case of the black-hole black-string transition there
are two overlapping black-string solutions but the non-uniform
one is unstable \cite{Gubser} and so the ``generalized
uniqueness'' holds. Note that {\it if} the latter is indeed
correct that would constitute another argument against the
existence of the stable non-uniform string of
\cite{HorowitzMaeda}, at least in the range of parameters where
the uniform string is stable.

For the rotating ring there are again three solutions in the
overlap region. Suppose we start with the black hole solution at
$J=J_1$, trace out the solution line (for fixed $M$), and denote
by region I the black hole solutions with $J$ going from $J_1$ to
$J_2$, by region II the black ring solutions going from $J_2$ to
$J_1$ and finally by region III the black ring going from $J_1$
to $J_2$. Since a stability analysis is not available at this
time, it is possible to provide a {\it concrete test} for the
stability criterion -- namely that the black ring solutions in
region II (on the right in figure \ref{BlackRing}) are unstable
(actually it would be enough if regions II and III have
instability domains which together cover the region $J_1 \le J
\le J_2$.

But why is there any reason to believe that the stability
criterion would be correct? The main tool so far in proving
uniqueness has been the positive energy theorems which may be
thought to be a strong generalization of the Green identity for
the Laplace equation, which is used to prove uniqueness for the
boundary value problem (either Dirichlet or Neumann). However,
this tool makes no reference to stability and so it must fail in
cases where several solutions exist. A possible direction is to
consider the phase transitions in the system. If there were two
stable phases with the same horizon topology then when one of
them would decay to the other, it could do so without a topology
change and so presumably without going through an explosive
singularity \cite{BKTopologyChange,BKExplosive}. It is
conceivable that such a mild phase transition is forbidden by
some ``censorship''.

\vskip 2cm
 \noindent {\em Acknowledgments}

It is a pleasure to thank G. Gibbons for comments.

Work supported by DOE under grant no. DE-FG02-90ER40542, and by a
Raymond and Beverly Sackler Fellowship.

\end{document}